# Postquantum eavesdropping without superluminal signaling

Antoni Wójcik*

Faculty of Physics, Adam Mickiewicz University,

Umultowska 85, 61-614 Poznań, Poland

In a recent Letter [1] Barrett, Hardy and Kent (BHK) considered a very interesting question which of the fundamental laws of physics ensure security of quantum cryptographic protocols. In particular, they presented quantum key distribution protocol and claimed its security in a very general framework, which allows - what they call - postquantum eavesdropping strategies i.e. which "allows for eavesdroppers who can break the laws of quantum mechanics, as long as nothing they can do implies the possibility of superluminal signaling". In this Comment, however, we show that it is still possible to construct a strategy, which although consistent with all asumptions explicitely stated in [1], enables perfect eavesdropping in BHK protocol.

In addition to no signaling condition (assumption A), BHK make only two further technical assumptions which restrict the possible action of Eve. Assumption B states that "measurements on a shared state cannot be used to send signals between the parties in any configuration" [1]. Finally, assumption C requires that "no information about events in Alice's and Bob's laboratories - in particular, their measurements or outcomes - subsequently propagates to Eve" [1]. The assumptions B and C mean that both Alice's and Bob's laboratories are secure against leaking information out. Note that none of these assumptions protect both laboratories against information flowing in the oposite direction.

Let us now present a scheme which shows that the assumptions A, B and C are not sufficient to ensure the security of the BHK protocol. Our attack concentrates on the second point of the BHK protocol. To choose measurement bases in a proper way, Alice and Bob have to use some sort of random number generator. It can be of classical or, better yet, of quantum nature. Let us now imagine that in a hypothetical postquantum world (far, far away) there exists Eve which has some capabilities of Jedi Masters [2]. Namely Eve is able to control results of random processes (e.g. quantum measurements). Let us also assume that the interaction between Eve and a given random number generator is mediated by some kind of postquantum field which propagates at a speed not exceeding the speed of light. Thus Eve cannot use her Force for superluminal signalling. Obviously, assumptions B and C hold as well. On the other hand, Eve can use Force to decide which measurements bases will be chosen by Alice and Bob. Having full control over the measurements bases Eve can also control the measurements results by sending to both Alice and Bob qubits in a deterministically prepared pure states corresponding to the bases chosen. In this way Eve knows perfectly the outcomes of all measurements without any information gained from Alice's and Bob' laboratories. Moreover, Eve can easily choose both measurements bases and their outcomes in such a way that they pass any desired correlation test (e.g. violation of some generalized Bell inequality).

We believe, that the above scheme does not impair the main conclusion of the BHK paper about the fundamental position of no signaling condition in quantum cryptography. It shows, however, that the BHK protocol in its actual version, although secure against particular attack analyzed in the Letter [1], is not secure against "general attacks by a postquantum eavesdropper limited only by the impossibility of superluminal signaling".

We would like to thank the State Committee for Scientific Research for financial support under grant no. 0 T00A 003 23.

*Email address: antwoj@amu.edu.pl

References

[1] J. Barrett, L. Hardy and A. Kent, Phys. Rev. Lett. 95, 010503 (2005)

[2] When the future of Anakin Skywalker is to be decided by a result of the chance cube tossing, Jedi Master Qui-Gon Jinn uses Force to win the game (*Star Wars Episode I: The Phantom Menace*).